# Imaging and spectroscopy of artificial-atom states in core/shell nanocrystal quantum dots

Oded Millo[1], David Katz[1], YunWei Cao[2], and Uri Banin[2*]

[1]Racah Institute of Physics
[2]Department of Physical Chemistry and the Farkas Center for Light Induced Processes,
The Hebrew University, Jerusalem 91904, Israel

## ABSTRACT

Current imaging scanning tunneling microscopy is used to observe the electronic wavefunctions in InAs/ZnSe core/shell nanocrystals. Images taken at a bias corresponding to the *s* conduction band state show that it is localized in the central core region, while images at higher bias probing the *p* state reveal that it extends to the shell. This is supported by optical and tunneling spectroscopy data demonstrating that the *s-p* gap closes upon shell growth. Shapes of the current images resemble atom-like envelope wavefunctions of the quantum dot calculated within a particle in a box model.

Semiconductor quantum dots (QDs), which represent the zero dimensional limit of the solid state, exhibit a set of discrete and narrow energy levels, and are therefore often considered as artificial atoms [1-4]. This elegant analogy, borne out from optical and tunneling spectroscopy [5], can be tested directly by observing the shapes of the QD electronic wavefunctions. Recently, the probability density of the ground and first excited states for epitaxially grown InAs QDs embedded in GaAs, was directly probed using scanning tunneling microscopy (STM) [6]. Magnetotunneling spectroscopy with inversion of the frequency domain data was also used to probe the spatial profiles of states of such QDs [7]. A different approach to QDs, with significant impact for fundamental studies, as well as for potential applications in nanotechnology, is the colloidal growth of free standing nanocrystals [1,8,9]. Unlike the MBE grown QDs, these nanocrystals have a nearly spherical shape leading to atom-like symmetries and degeneracies for the electronic states [5]. For such QDs, we use the unique sensitivity of the STM to the electronic density of states on the nanometer scale [6,10,11], to directly image the *s* and *p*-like wavefunctions. The interpretation of the images as manifesting the wavefunctions is supported by optical and tunneling spectra acquired on these QDs. These combined measurements yield unique information on the energetics, spatial extent and shapes of the QD states, providing a test of the theoretical understanding of QDs.

We investigate novel InAs/ZnSe core/shell nanocrystals [12]. The core/shell configuration, where a shell of a high band-gap semiconductor is overgrown on the core, provides further control over the electronic and optical properties of nanocrystals, in addition to size [13,14]. The QDs were prepared using a two-step high temperature solution-phase synthesis of organometallic precursors. First, InAs cores 1.7 nm in radius were prepared and purified from the growth solution resulting in a size distribution of better than 10%. At the second stage ZnSe shells up to six monolayers (ML) thick, with ~1ML accuracy, were overgrown on these cores [12]. The fluorescence quantum yield was significantly enhanced upon shell growth from an absolute value of 1% for the cores, up to 20% for an optimal shell thickness of around 1.5 ML [12]. The outer shell surface is passivated by organic ligands, and for the cryogenic STM measurements, all performed at 4.2 K, we link the nanocrystals to a gold film via hexane dithiol molecules (DT) [5]. The tunneling spectra were measured after positioning the STM tip above the QD, forming a double barrier tunnel junction (DBTJ) as detailed in Ref. 5.

Fig. 1 shows tunneling-conductance spectra, dI/dV vs. V curves, measured on two core/shell nanocrystals with 2 and 6 ML shells, along with a typical curve for an InAs QD of radius similar to the nominal core radius ~1.7 nm. We either numerically differentiate the I-V curves or use a direct lock-in method, both yielding similar results. We employed an asymmetric configuration of the DBTJ, so that ~90% of the applied voltage drops on the tip-QD junction. This was achieved by retracting the tip while measuring the I-V characteristics until no apparent change in the peak spacings was observed, typically obtained at set-bias of 2V and set current of 0.1 nA. Consequently, the conduction (valence) band states appear at the positive (negative) bias side, and the QD level spacings can be directly extracted from the peak spacings [5,15]. The spectra of core and core/shell nanocrystals exhibit single electron tunneling peaks, and their general appearance is similar. The gap in the density of states around zero bias, associated with the QD band-gap, is nearly identical. On the positive bias side a doublet of peaks is observed, followed by a higher order multiplet. As discussed previously for the cores [5], the doublet corresponds to tunneling through the two-fold degenerate $s$-like conduction band (CB) state (denoted $1S_e$ [9]), with the spacing assigned to the single electron charging energy. The higher order multiplet is assigned to tunneling through the $p$-like state ($1P_e$), reflecting the atomic-like nature of the QD levels.

Nevertheless, distinct differences are observed between the core and core/shell tunneling spectra. While the band-gap is nearly unaffected by the shell growth, the $s$-$p$ level separation is substantially reduced. Both effects are consistent with a model in which the $s$ state is confined to the InAs core region, while the $p$ level extends to the ZnSe shell. In this case, the $p$ state is red-shifted upon increasing shell thickness, whereas the $s$ level does not shift, yielding a closure of the CB $s$-$p$ gap. Due to the large valence band (VB) offset between InAs and ZnSe, the VB edge state also remains intact. Therefore, the band-gap is nearly unaffected by the shell growth, as observed also in the optical absorption spectra [12].

The different extent of the CB $s$ and $p$ states, implied by our spectroscopic results, can be directly probed by using the STM to image the QD atomic-like wavefunctions. To this end, we performed bias-dependent current imaging measurements [6], and a typical experiment on a core/shell nanocrystal with 6 ML shell is shown in Fig. 2. First, a dI/dV - V spectrum was acquired (Fig. 2(a)), and the bias for tunneling to the $s$ and $p$ states were determined. Then, a topographic image was measured at a bias value above the $s$ and $p$ states, $V_B$ = 2.1 V (Fig. 2(b)), simultaneously with 3 current images. At each point along the topography scan the STM feedback circuit was disconnected momentarily, and the current was measured at three different $V_B$ values: 0.9 V - corresponding to the CB $s$ state (Fig. 2(c)), 1.4 V - within the $p$ multiplet (2(d)), and 1.9 V - above the $p$ multiplet (2(e)). Therefore, the topographic and current images are all measured with the same constant local tip-QD separation. The main factor determining each current image is thus the local (bias-dependent) density of states, reflecting the shape of the QD electronic wavefunctions.

Upon comparing the current images, pronounced differences are observed in the extent and shape of the $s$ and $p$ wavefunctions. The image corresponding to the $s$-like wavefunction (Fig. 2(c)) is localized to the central region of the core/shell nanocrystal, while the images corresponding to the $p$-like wavefunctions extend out to the shell (Figs. 2(d-e)), consistent with the model discussed above. This can also be seen in the cross-sections presented in frame ($f_1$) taken along a common line through the center of each current image, and most clearly in ($f_2$), which shows the current normalized to its maximum value along the same cuts. Image 2(e), taken at a voltage above the $p$ multiplet, manifests a nearly spherical geometry similar to that of image 2(c) for the $s$ state but has a larger spatial extent. Image 2(d), taken with $V_B$ near the middle of the $p$ multiplet, is also extended but has a truncated top with a small dent in its central region.

To aid the interpretation of the current images, we calculated the envelope wavefunctions for the CB states using an illustrative model assuming a spherical shape for the QD with the radial potential taken as shown in the inset of Fig. 2(a) [12,16]. The energy calculated for the $s$ state is lower than the barrier height at the core-shell interface, and has the same values for core and core/shell QDs. In contrast, the energy of the $p$ state is above the core-shell barrier and it red shifts with shell growth, in qualitative agreement with our spectroscopic results. Isoprobability surfaces for the different wavefunctions are presented in Figs. 2(g-i), with Fig. 2(g) showing the $s$ state, Fig. 2(h) the in-plane component of the $p$ wavefunctions, $p_x^2+p_y^2$, that has a torus-like shape, and Fig. 2(i) depicting the two lobes of the perpendicular component, $p_z^2$. The square of the radial parts of the $s$ and $p$ wavefunctions are presented in Fig. 2(j). The calculated probability density for the $s$ state is spherical in shape and mostly localized in the core, consistent with the experimental image taken at a bias where only this level is probed (Fig. 2(c)). The $p$ components extend much further to the shell as observed in the experimental images taken at higher bias. Moreover, the different shapes observed in the current images can be assigned to different combinations of the probability density of the $p$ components.

A filled torus shape, similar to the current image 2(d) taken at the middle of the $p$ multiplet, can be obtained by a combination with larger weight of the in-plane $p$ component ($p_x^2+p_y^2$), parallel to the gold substrate, and a smaller contribution of the perpendicular $p_z$ component. The non-equal weights reflect preferential tunneling through the in-plane components. This may result from a perturbation due to the specific geometry of the STM experiment leading to a small degeneracy lifting. A spherical shape for the isoprobability surfaces results from summing all the $p$ components with equal weights, consistent with the current image measured at a bias above the $p$ manifold (Fig. 2(e)).

Current images obtained at bias values corresponding to the $p$ states, have a contribution also from tunneling through the $s$ state. However, this contribution is small in InAs/ZnSe core/shell nanocrystals having a thick shell, due to the reduced tunneling probability through the $s$ wavefunction, which is localized in the core. Therefore, our choice of such a nanocrystal system provides a suitable contrast mechanism for STM imaging of the different QD wavefunctions, allowing for a direct visualization of their atomic-like character.

In similar experiments performed on other InAs/ZnSe nanocrystals (total of 10 QDs), we always observed that the $s$ state is far more localized than the $p$ states. Furthermore, the shapes of current images obtained at bias values corresponding to the $s$ state, as well as above the $p$ manifold, lack nodal signature. The current images obtained at intermediate bias values for 7 particles exhibited a clear torus-like shape. However, in some cases other shapes were seen, corresponding to different combinations of the $p$ wavefunction components. For example, a double lobe structure resembling a single in-plane component was observed in two cases, manifesting an even stronger degeneracy lifting induced by the local QD environment.

Optical spectroscopy provides further evidence for the picture borne out from our tunneling data. In particular, the photoluminescence excitation (PLE) spectra presented in Fig. 3 manifest the control over the QD level structure afforded by the shell growth, leading to the reduction of the *s-p* level spacing. The three spectra, for cores (solid line), and core/shells with 4 ML and 6 ML shell thickness (dotted and dashed lines, respectively), were measured at T=10 K using the same detection window (970 nm), corresponding to the excitonic band-gap energy for InAs cores 1.7 nm in radius [17]. These spectra were measured at low excitation powers, using a monochromatized tungsten lamp. The peak labeled III, which corresponds in the cores to the transition from the VB edge state to the CB *p* state [5], is red shifted monotonically upon shell growth.

The dependence of the difference between peak III and the band-gap transition I on shell thickness is depicted in the inset of Fig. 3 (circles), along with the *s-p* gap extracted from the tunneling spectra by subtracting the charging energy from the inter-multiplet spacing (squares). The qualitative trend of red shift is similar for both data sets, but there is a quantitative difference with the optical shift being considerably smaller. This is in contrast to the good correlation between the optical and tunneling spectra observed previously for InAs cores [5], providing an opportunity to examine the intricate differences between these two complementary methods. First, while the tunneling data directly depict the spacing between the two CB states, the PLE data in the inset of Fig. 3 represent the energy difference between two VB to CB optical transitions. Therefore, a blue shift of the *p*-like component [18] of the VB edge state upon shell growth will reduce the net PLE red shift of transition III, as compared to the *s-p* gap depicted from the tunneling data. A modification in the complex VB level structure of the core/shells is also reflected in additional changes of the PLE spectra upon shell growth, e.g., transition II in the PLE (assigned in the cores to $2S_{3/2}$-$1S_e$ [17]) looses oscillator strength. Furthermore, in the tunneling spectra, the negative bias side tentatively assigned to the VB, shows a more complex peak structure as compared to the core. A second factor that may contribute to the deviation between the tunneling and optical *s-p* closure is the presence of charging effects only in the tunneling experiment. However, in our previous experiments on InAs cores [5], it appeared that e-e interactions did not have a considerable effect on the neutral QD level structure, which is measured in the optical experiment.

In conclusion, we have imaged the envelope wavefunctions confined in core/shell nanocrystal quantum dots using STM. The images portray the *s* and *p*-like wavefunctions and show that they differ considerably in their spatial extent. This is reflected also in our tunneling and optical spectra, both showing a closure of the *s-p* gap upon shell growth. This combination of wavefunction imaging and spectroscopic techniques allowed us to visualize the atomic-like character of nanocrystal quantum-dots.

This work was supported by the Israel Science Foundation and the BIKURA program.

**FIGURE CAPTIONS:**

**FIG. 1:** Tunneling conductance spectra of an InAs core QD (numerical derivative) and two core/shell nanocrystals with two (numerical derivative) and six (lock-in method) ML shells with nominal core radii ~ 1.7 nm. The spectra were offset along the V direction to center the observed zero current gaps at zero bias.

**FIG. 2:** Wavefunction imaging and calculation for an InAs/ZnSe core/shell QD having a six ML shell. (a) A tunneling spectrum (lock-in method) acquired for the nanocrystal. (b) 8x8 $nm^2$ topographic image taken at $V_B$ = 2.1 V and $I_s$=0.1 nA. (c-e) Current images obtained

simultaneously with the topographic scan at three different bias values denoted by arrows in (a). A 3x3 median filter was applied. ($f_1$) Crossections taken along the diagonal of the current images at 0.9 V (magenta), 1.4 V (blue) and 1.9 V (green). ($f_2$) The same crossections normalized to their maximum current values. (g-j) Envelope wavefunctions calculated within a 'particle in a sphere' model. The radial potential and the energies of the *s* and *p* states are illustrated in the inset of frame (a). (g-i) Isoprobability surfaces, showing $s^2$ (g), $p_x^2+p_y^2$ (h), and $p_z^2$ (i). (j) The square of the radial parts of the *s* and *p* wavefunctions normalized to their maximum values. For the core-shell potential offset we used the bulk InAs-ZnSe value, 1.26 eV. The potential offset at the shell surface was taken as 8 eV. Bulk InAs and ZnSe electron effective masses were used ($m_e^*$=0.024 and 0.13, respectively), and in the matrix region we used the free electron mass.

**FIG. 3:** PLE spectra, normalized to peak III, for InAs cores (solid line) and InAs/ZnSe core/shell nanocrystals of four (dotted line) and six (dashed line) ML shells, with the zero of the energy scale taken at the detection window (970 nm). The inset depicts the dependence on shell thickness of the *s-p* level closure, as determined by tunneling (open squares) and PLE (solid circles). The error bars in the tunneling data represent the minimum to maximum spread in the *s-p* spacings measured on 5-10 QDs for each sample, most likely arising from the distribution in core radii and shell thickness. The PLE data points represent the difference between transition III and transition I (transition I, which hardly shifts upon shell growth, is taken as the central point between the detection energy and the energy of the first PLE peak [17]), averaged over three detection windows (950, 970 and 990 nm).

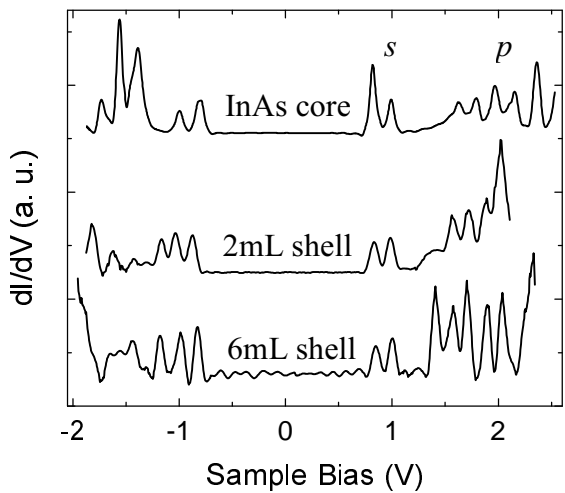

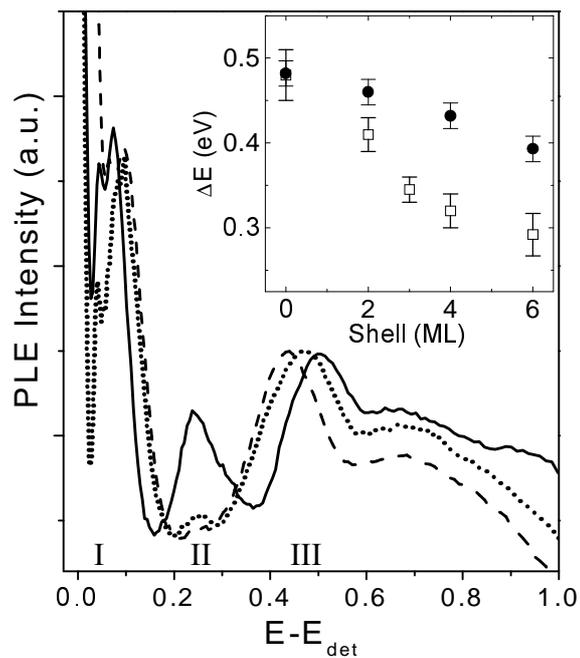

*Figure 1, Millo et al.*

*Figure 3, Millo et al.*

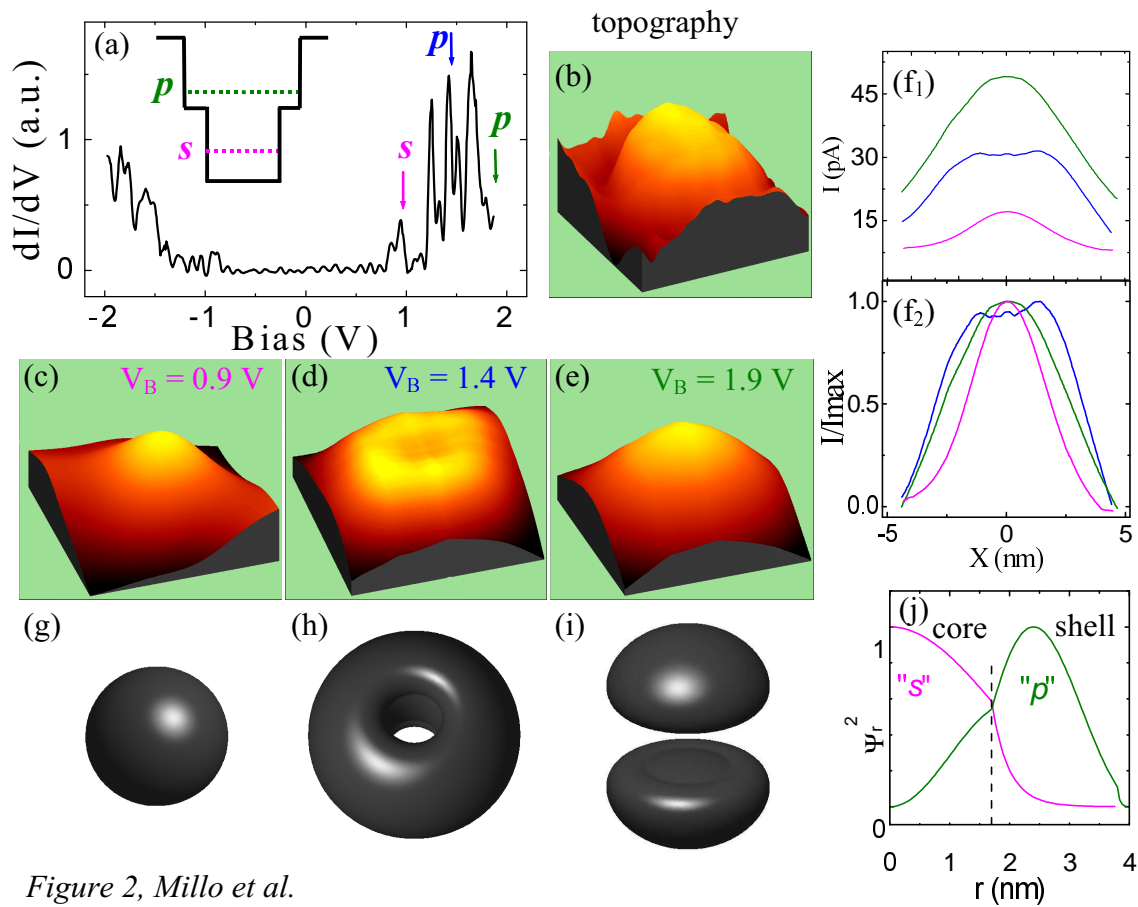

*Figure 2, Millo et al.*